\documentclass[twocolumn,amsmath,amssymb,aps]{revtex4-1}
\usepackage{graphicx}
\usepackage{dcolumn}
\usepackage{bm}
\usepackage{algorithm}
\begin{document}


\title{Shallow-circuit variational quantum eigensolver based on symmetry-inspired Hilbert space partitioning for quantum chemical calculations}
\author{Feng Zhang}
 \email{fzhang@ameslab.gov}
 \affiliation{Ames Laboratory, US Department of Energy, Ames, IA 50011, USA.}
 \author{Niladri Gomes}
 \affiliation{Ames Laboratory, US Department of Energy, Ames, IA 50011, USA.}
\author{Noah F. Berthusen}
 \affiliation{Ames Laboratory, US Department of Energy, Ames, IA 50011, USA.}%
 \affiliation{Department of Physics and Astronomy, Iowa State University, Ames, IA 50011, USA.} 
\author{Peter P. Orth}
 \affiliation{Ames Laboratory, US Department of Energy, Ames, IA 50011, USA.}%
 \affiliation{Department of Physics and Astronomy, Iowa State University, Ames, IA 50011, USA.}
 \author{Cai-Zhuang Wang}
  \affiliation{Ames Laboratory, US Department of Energy, Ames, IA 50011, USA.}%
 \affiliation{Department of Physics and Astronomy, Iowa State University, Ames, IA 50011, USA.}%
 \author{Kai-Ming Ho}%
 \affiliation{Ames Laboratory, US Department of Energy, Ames, IA 50011, USA.}%
 \affiliation{Department of Physics and Astronomy, Iowa State University, Ames, IA 50011, USA.}
 \author{Yong-Xin Yao}
 \email{ykent@iastate.edu}
 \affiliation{Ames Laboratory, US Department of Energy, Ames, IA 50011, USA.}%
 \affiliation{Department of Physics and Astronomy, Iowa State University, Ames, IA 50011, USA.}
\date{\today}
\begin{abstract}
Development of resource-friendly quantum algorithms remains highly desirable for noisy intermediate-scale quantum computing. Based on the variational quantum eigensolver (VQE) with unitary coupled cluster ansatz, we demonstrate that partitioning of the Hilbert space made possible by the point group symmetry of the molecular systems greatly reduces the number of variational operators by confining the variational search within a subspace. In addition, we found that instead of including all subterms for each excitation operator, a single-term representation suffices to reach required accuracy for various molecules tested, resulting in an additional shortening of the quantum circuit. With these strategies, VQE calculations on a noiseless quantum simulator achieve energies within a few meVs of those obtained with the full UCCSD ansatz for $\mathrm{H}_4$ square, $\mathrm{H}_4$ chain and $\mathrm{H}_6$ hexagon molecules; while the number of controlled-NOT (CNOT) gates, a measure of the quantum-circuit depth, is reduced by a factor of as large as 35. Furthermore, we introduced an efficient ``score" parameter to rank the excitation operators, so that the operators causing larger energy reduction can be applied first. Using $\mathrm{H}_4$ square and $\mathrm{H}_4$ chain as examples, We demonstrated on noisy quantum simulators that the first few variational operators can bring the energy within the chemical accuracy, while additional operators do not improve the energy since the accumulative noise outweighs the gain from the expansion of the variational ansatz.    
\end{abstract}
\maketitle
\section{Introduction}
Quantum computers have been projected to be the ultimate solution to classically intractable problems owing to the exponential expansion of information that can be processed on quantum bits (qubits) compared with classical bits. However, to fully realize the advantage of quantum computing, quantum devices that integrate thousands or more qubits with sufficiently long coherent time have to be developed, which remains a significant challenge as of today. In the foreseeable future, one still has to work with so-called noisy intermediate-scale quantum devices (NISQ)~\cite{Preskill2018}, and practical quantum algorithms need to be made aware of this limitation. One group of such algorithms is the variational quantum eigensolver (VQE)~\cite{Peruzzo2014}, which uses a variational approach to optimize an objective function on a quantum/classical hybrid architecture. The preparation of parametrized quantum states and measurement of the expectation value of the objective function are performed on a quantum computer with relatively shallow circuits, while an optimization algorithm is implemented on a classical computer to find the optimal parameters. 

Quantum chemistry has been one of the most active fields for quantum computing~\cite{McArdle2020}, realizing a proposal of solving quantum-chemical problems on a quantum architecture that Feynman made nearly 30 years ago ~\cite{Feynman1982}. There has been significant development of using VQE to solve quantum-chemical problems in both theory~\cite{Wecker2015, McClean2016, OMalley2016, McClean2017, Barkoutsos2018, Colless2018, Romero2019, Higgott2019, Lee2019, Grimsley2019, Tang2019} and experiments on real quantum devices~\cite{Peruzzo2014,OMalley2016,Colless2018,Kandala2017,Hempel2018,Shen2017,Yao2020}. The most commonly used ansatz is derived from the unitary coupled cluster (UCC) method~\cite{Hoffmann1988, Bartlett1989, Romero2019}, which is an extension of the well-known coupled cluster theory for describing the correlation effects in quantum systems~\cite{Helgaker2014}. In most applications, only single and double excitations are included, resulting in UCCSD. With this truncation, UCCSD in general cannot reach the true ground state energy. Another necessary step for implementing UCCSD is Trotterization~\cite{Trotter1959}, that is, the expansion of $e^{A+B}$ as $\left(e^{A/n}e^{B/n}\right)^n$, where $e^{A/n}$ and $e^{B/n}$ can be efficiently implemented on quantum computers using available one- and two-qubit gates~\cite{McArdle2020}. This expansion is exact only in the limit of $n\to\infty$ when operators $A$ and $B$ do not commute. Conventionally, only a single Trotter step ($n=1$) was used to represent UCCSD; more trotter steps barely improve the ansatz while significantly elongate the quantum circults~\cite{OMalley2016,Hempel2018}. A variation of UCCSD was also introduced recently, which, by successively adding operators to the ansatz one at a time in an adaptive process, can reach the accuracy of the true ground state with relatively shallow circults~\cite{Grimsley2019}.

Despite the truncation and Trotterization, the implementation of UCCSD VQE on real devices has still been limited to small molecules, including $\mathrm{H}_2$, $\mathrm{HHe}^+$, $\mathrm{LiH}$ and $\mathrm{BeH}_2$~\cite{Peruzzo2014,OMalley2016,Colless2018,Kandala2017,Hempel2018,Shen2017}. We note that VQE has also been applied to effective interacting few-site models that emerge from infinite lattice systems within Gutzwiller embedding theory~\cite{Yao2020, ga_pu, Lanata2017}. In this paper, we use the intrinsic symmetry to further simplify UCCSD to meet NISQ requirements. The implementation of symmetry in constructing variational ans\"atze is not a new concept. In fact, the particle-number symmetry and $Z_2$ symmetry have been fully considered in selecting the excitation operators in UCCSD~\cite{Romero2019,Grimsley2019}. The point group symmetry of a molecule can also prohibit certain spin-orbital excitations~\cite{Hempel2018, Fischer2019}. More specifically, the point-group symmetry can further divide the Hilbert space preserving quantum numbers associated with the particle-number and $Z_2$ symmetries into several subspaces. Here, we will introduce an efficient graph clustering technique to identify these subspaces in the qubit representation. This general scheme allows us to systematically identify the most relevant excitation operators for the purpose of constructing a trial state with an energy close to the ground state energy. Based on the Hamiltonian matrix, this method is numerically cheap and does not require a sophisticated group theoretical analysis of the problem. 
It can be further combined with other strategies to reduce the gate complexity of the resulting circuit. Below, we will establish an importance ordering among the different excitation operators and combine all subterms associated with a particular operator. The resulting circuits are significantly shortened, allowing us to efficiently reach chemical accuracy for molecules $\mathrm{H}_4$ and $\mathrm{H}_6$. The calculations were performed using the toolkit QISKiT developed by IBM~\footnote{python api. https://qiskit.org/}, with both noise-free statevector and noisy QASM simulators.

\section{Formalism and results}
The second quantization is applied to construct the Hamiltonian of the molecules. Atomic orbitals in the minimal basis (STO-3g)~\cite{Hehre1969} are used in the calculations. Relevant spin-orbitals for constructing the basis of the Hilbert space are determined according to the Hartree-Fock (HF) calculations. In the second-quantized formulation, the electronic Hamiltonian is expressed as 
\begin{equation}\label{sq}
    H=\sum_{pq,\sigma}h_{pq}a^\dagger_{p\sigma}a_{q\sigma}+\frac{1}{2}\sum_{pqrs,\sigma\lambda}h_{pqrs}a^\dagger_{p\sigma}a^\dagger_{q\lambda}a_{r\lambda}a_{s\sigma},
\end{equation}
where $h_{pq}$ and $h_{pqrs}$ are one-electron and two-electron integrals, respectively, and $\sigma$ and $\lambda$ denote spins. $h_{pq}$ and $h_{pqrs}$ are calculated with the PySCF package~\cite{PYSCF}. The creation and annihilation operators in Eq.~\ref{sq} are defined on $2N$ spin-orbitals ($N$ is the total number of electrons).

In order to solve the Hamiltonian on a qubit-based quantum computer, it is necessary to transform the Fock state $\lvert f_{2N},f_{2N-1},\cdots,f_1\rangle$, where $f_i$ is the occupation number of the $i^{th}$ spin-orbital (0 or 1), to a qubit state $\lvert q_{2M},q_{2N-1},\cdots,q_1\rangle$ with $M\leq N$. Accordingly, the ferminoic operators in Eq.~\ref{sq} are transformed to qubit operators that can be realized in quantum circuits based on Pauli gates. We use the parity encoding method~\cite{Seeley2012}, in which the $p^{th}$ qubit stores the parity of the total occupation number of the first $p$ spin-orbitals: $q_p=\left[\sum_{i=1}^pf_i\right]$ (mod 2). If the spin-orbitals in the Fock state are arranged in such a way that the first $N$ spin-orbitals describe spin-up states and the last $N$ spin-orbitals describe spin-down states, then $q_N$ and $q_{2N}$ are equal to the number of spin-up electrons (mod 2) and the number of electrons (mod 2), respectively. For non-relativistic molecules, these two numbers will be conserved. Consequently, the two qubits $q_N$ and $q_{2N}$ will only be acted on by identity or Pauli $Z$ operators, which can then be replaced by the corresponding eigenvalues, resulting in a Hamiltonian that only acts on $2M=2N-2$ qubits~\cite{McArdle2020}. In other words, with the parity encoding method, one can effectively save two qubits in quantum computing.  

VQE employs the Rayleigh-Ritz variational principle
\begin{equation}\label{rr}
    \frac{\langle\psi(\vec{\theta})|H|\psi(\vec{\theta})\rangle}{\langle\psi(\vec{\theta})|\psi(\vec{\theta})\rangle}\geq E_0,
\end{equation}
where $E_0$ is the ground state energy and $\vec{\theta}$ are variational parameters. A quantum circuit with moderate depth is used to apply a unitary operator $U(\vec{\theta})$ on the initial state $\lvert\psi_0\rangle$, which is chosen to be the HF state in our calculations, creating a parametrized state $\lvert\psi(\vec{\theta})\rangle$: $\lvert\psi(\vec{\theta})\rangle=U(\vec{\theta})\lvert\psi_0\rangle$. The expectation value  $\langle\psi(\vec{\theta})|H|\psi(\vec{\theta})\rangle$ is measured on a quantum computer, while $\vec{\theta}$ are varied to minimize the Rayleigh-Ritz quotient (left hand side of Eq.~\ref{rr}) on a classical computer. Unitary coupled cluster (UCC) is a chemistry-inspired ansatz that has been widely used in solving quantum-chemical problems~\cite{Helgaker2014}. In UCC, $U(\vec{\theta})$ can be written as $U(\theta)=e^{\theta(T-T^\dagger)}$, where $T$ can be any Hermitian excitation operator. However, only single and double excitations are usually selected, and UCC in this form is called UCCSD:
\begin{equation}
    T=\sum_{i\alpha}\theta_{i\alpha}a^\dagger_ia_\alpha+\sum_{ij\alpha\beta}\theta_{ij\alpha\beta}a^\dagger_ia^\dagger_ja_\alpha a_\beta,
\end{equation}
where the subscripts $\alpha\beta$ and $ij$ denote occupied and virtual spin-orbitals, respectively. Using a single Trotter step, the UCCSD ansatz can be expressed as
\begin{equation}\label{trotter}
    U(\vec{\theta})=\prod_{i\alpha}e^{\theta_{i\alpha}(a^\dagger_ia_\alpha-a^\dagger_\alpha a_i)}\prod_{ij\alpha\beta}e^{\theta_{ij\alpha\beta}(a^\dagger_ia^\dagger_ja_\alpha a_\beta-a^\dagger_\beta a^\dagger_\alpha a_ja_i)}.
\end{equation}
The occupied and virtual spin-orbitals in Eq.~\ref{trotter} are selected in such a way that the net magnetization of the molecule is conserved. The total number of excitation operators in UCCSD for a non-magnetic system with $N$ electrons can be calculated as: $M=2(N/2)^2+N/2(N/2-1)+(N/2)^4$. $M$ increases rapidly with $N$, making it challenging to implement the UCCSD ansatz for even moderate values of $N$ on NISQ. For instance, $M=26$ when $N=4$, which already requires over a thousand CNOT gates to prepare the ansatz state $\psi(\vec{\theta})$.

An $n$-qubit system can span a Hilbert space with a dimension of $2^n$. By construction, all the excitation operators included in UCCSD only act on a subspace preserving the total number of electrons and the net magnetization. This subspace separates into smaller subspaces if there are additional symmetry elements from the structure of the molecule. The ground-state of the Hamiltonian lies in one of these subspaces. Once this subspace is identified, one can confine the variational search within this subspace. That is, starting from an initial state $|\psi_0\rangle$ in this subspace, one only needs to apply the excitation operators that keep the ansatz states $|\psi(\vec{\theta})\rangle$ in the same subspace. In this way, the number of excitation operators in the variational ansatz can be greatly reduced. 

Let us analyze the ${\mathrm H}_2$ dimer as an example to illustrate this idea~\cite{OMalley2016}. With two-qubit reduction applied in the parity encoding scheme, the fermionic system can be mapped onto a 2-qubit system, spanning a 4-dimensional Hilbert space. The four basis vectors in this qubit space $|00\rangle$, $|01\rangle$, $|10\rangle$ and $|11\rangle$ (parity encoding) correspond to the four Fock states $|f_{2\downarrow}, f_{1\downarrow}, f_{2\uparrow}, f_{1\uparrow}\rangle =|0110\rangle$, $|0101\rangle$, $|1010\rangle$ and $|1001\rangle$, respectively. All the four Fock states conserve the total number of electrons (2) and the net spin (0). At a H-H distance of 0.725 \AA, the Hamiltonian in Eq.~\ref{sq} can be represented by the following $4\times4$ matrix (in units of eV):
\begin{equation}\label{hamilH2}
    H=\begin{pmatrix}
       -1.06 & 0 & 0 & 0.18 \\
       0 & -1.84 & 0.18 & 0 \\
       0 & 0.18 & -0.23 & 0 \\
       0.18 & 0 & 0 & -1.06
       \end{pmatrix}
\end{equation}
By inspection, it is easily seen that the 4-dimensional Hilbert space can be separated into two subspaces $S_1$ and $S_2$, spanned by \{$|00\rangle$, $|11\rangle$\} and \{$|01\rangle$, $|10\rangle$\}, respectively. The HF state is represented by the qubit state $|01\rangle$ (or the Fock state $|0101\rangle$). Starting from $|01\rangle$, one needs to select excitation operators that flip both qubits so that the ansatz states remain in the same subspace. There are two single excitation operators and one double excitation operator in the full UCCSD for $\mathrm{H}_2$: $a^\dagger_2a_1-a^\dagger_1a_2$, $a^\dagger_4a_3-a^\dagger_3a_4$ and $a^\dagger_2a^\dagger_4a_3a_1-a^\dagger_1a^\dagger_3a_4a_2$, which are transformed to spin operators $iY_1$, $iY_2$, $\frac{i}{2}(X_2Y_1-Y_2X_1)$, respectively. Here, $X$ and $Y$ are Pauli matrices, and the subscripts specify which qubit the Pauli matrix acts on. The two single hopping terms $Y_1$ and $Y_2$ transfer $|01\rangle$ onto $|00\rangle$ and $|11\rangle$, respectively, both of which are out of the subspace where the initial state $|01\rangle$ is located. Therefore, the only relevant operator is the double excitation term $i/2(X_2Y_1-Y_2X_1)$.
 
\begin{figure*}
\includegraphics[width=\linewidth]{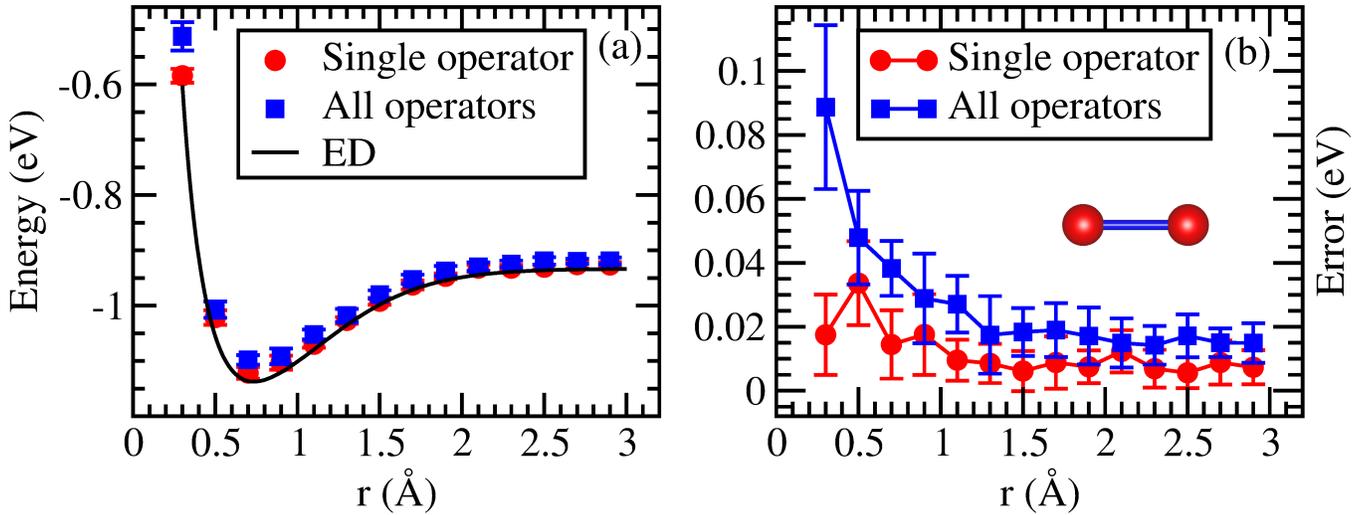}
\caption{\label{h2ground}(a) The ground state energy of $\mathrm{H}_2$ as a function of H-H distance, calculated with ED and on a two-qubit QASM simulator. (b) The error of QASM results as compared with ED values. The error bar was determined based on 10 independent simulations with 1024 shots in each simulation (the same below). The inset in (b) shows the geometric configuration of a H$_2$ dimer.}
\end{figure*}

There is another simplification that can be made. $Y_2X_1=X_2Y_1Z_2Z_1$ using the relation $\sigma_i\sigma_j=\delta_{ij}I+i\epsilon_{ijk}\sigma_k$, where $\sigma_1$, $\sigma_2$, $\sigma_3$ stands for $X$, $Y$, $Z$, respectively, $I$ is the identity matrix, and $\epsilon_{ijk}$ is the parity of the permutation $(ijk)$. Any state in the subspace spanned by $|01\rangle$ and $|10\rangle$ is an eigenstate of $Z_2Z_1$ with an eigenvalue of -1. Thus, $Z_2Z_1$ can be replaced with -1 when acting on this subspace: $X_2Y_1Z_2Z_1=-X_2Y_1$. Consequently, the two terms in the double excitation operator $\frac{i}{2}(X_2Y_1-Y_2X_1)$ can be combined into one term $iX_2Y_1$, further reducing the circuit length by half. 

We performed VQE calculations on the ground-state energy of $\mathrm{H}_2$ molecule, using the QASM quantum simulator as implemented in the quantum computing toolkit QISKiT~\cite{Note1}. To simulate real NISQ devices, a noise model is implemented by including depolarizing gate errors for all qubits participating in the gate.
To investigate the effect of circuit simplification, we implemented two different variational ans\"atze $e^{i\theta_3/2(X_2Y_1-Y_2X_1)}e^{i\theta_2 Y_2}e^{i\theta_1 Y_1}$ and $e^{i\theta X_2Y_1}$, corresponding to the full UCCSD and its simplified form, respectively. Since UCCSD is exact for $\mathrm{H}_2$, both ans\"atze are expected to give the exact diagonalization (ED) results without noise. The fermionic Hamiltonian in Eq.~\ref{sq} is mapped onto a sum of tensor products of Pauli matrices (Pauli strings). The expectation value of each Pauli string was measured separately by averaging over 1024 shots. The error associated with the imperfect averaging is also included in the QASM simulator. The variational parameters are updated classically using the simultaneous perturbation stochastic approximation (SPSA) algorithm with 200 maximal iterations. The QASM simulations were performed 10 times independently to obtain an estimation of the error bar. In Fig.~\ref{h2ground} (a), we plot the dissociation curve for $\mathrm{H}_2$, calculated by QASM simulations and ED. In Fig.~\ref{h2ground} (b), we show the error for the two different ans\"atze. The simplified ansatz containing a single term $iX_2Y_1$ gives smaller error at all bond lengths. While noticeable fluctuations exist in different trials as can be seen from the error-bar size, the average values for the single-term ansatz give acceptable accuracy, with an averaged error of 11.7 meV over all all measured bond lengths. For the full UCCSD ansatz, the error bars are larger, and the averaged error is increased to 27.1 meV, clearly demonstrating that the shortened quantum circuit results in a significant noise reduction.

\begin{figure*}
\includegraphics[width=\linewidth,clip]{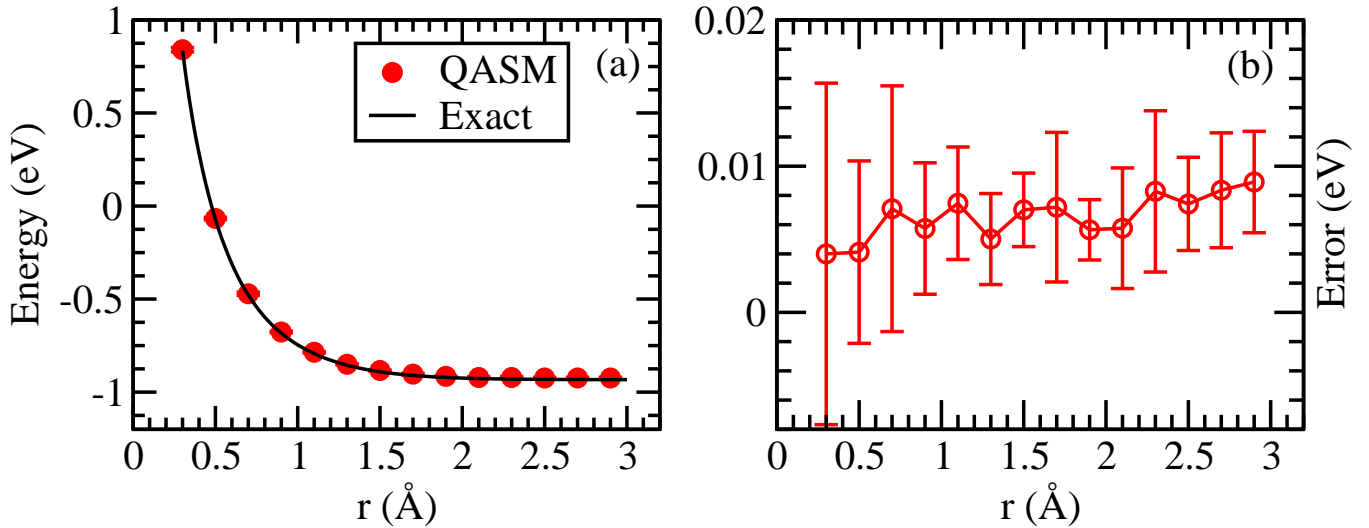}
\caption{\label{h2exi}(a) The first exited state energy of $\mathrm{H}_2$ as a function of H-H distance, calculated with ED and on a two-qubit QASM simulator. (b) The error of QASM results as compared with ED values.}
\end{figure*}

The first excited state of $\mathrm{H}_2$ lies in the subspace spanned by $|00\rangle$ and $|11\rangle$. Thus, one can also obtain the energy of the first excited state by implementing VQE in this subspace~\cite{Higgott2019,Lee2019}. We performed QASM simulations to verify this. Here, only a single term $iX_2Y_1$ is included in the variational ansatz, but we choose $|00\rangle$ as the initial state. In Fig.~\ref{h2exi} (a), we plot the first excitation state energy as a function of $r$, obtained from QASM simulations and ED; while in (b), we show the simulation error as a function of $r$. Again, accurate results can be obtained, with the averaged error being 6.6 meV. 

It is a non-trivial task to identify the Hilbert space separation, especially when the ferminoic Hamiltonian is transformed to a qubit representation.  We use the following algorithm based on graph clustering. A graph is created by denoting each basis vector of the Hilbert space as a node and connecting any two nodes $i$ and $j$ with an edge if the qubit Hamiltonian matrix element $H_{ij}$ is larger than a cutoff value: $\lvert H_{ij}\rvert>\epsilon_c$. Here, $\epsilon_c$ is set to $10^{-6}$ eV. The problem to be solved is to separate the total space of connected nodes into isolated clusters so that any two nodes within the same cluster can be linked with a continuous path (not necessarily directed connected with an edge), while such a path does not exist for any two nodes belonging to different clusters. We provide the algorithm for solving this clustering problem in pseudocodes in the Appendix.  Each cluster will then represent a separate subspace. To ensure the VQE calculation is confined in one of these subspaces, we use the following procedure to select the fermionic excitation operators. Assume the $m$ qubit basis vectors of the subspace correspond to $m$ Fock states: $\lvert\psi_0\rangle,\lvert\psi_1\rangle,\cdots$, and $\lvert\psi_{m-1}\rangle$. $\lvert\psi_0\rangle$ is pre-selected as the initial state for VQE calculations. Every other state $\psi_i$ ($0<i<m$) is related to $\lvert\psi_0\rangle$ by an excitation operator $T_i$: $\lvert\psi_i\rangle=T_i\lvert\psi_0\rangle$. The operators will be selected from all $U_i=T_i-T^\dagger_i$ (so that $e^{\theta U}$ is unitary), provided they satisfy the following two conditions:
\begin{itemize}
    \item $U_i$ contains only single or double excitation.
    \item $U_i$ does not transfer any of the basis vectors $|\psi_0 \rangle, \ldots, |\psi_{m-1} \rangle$ out of the subspace.
\end{itemize}

\begin{figure}
\includegraphics[width=\linewidth]{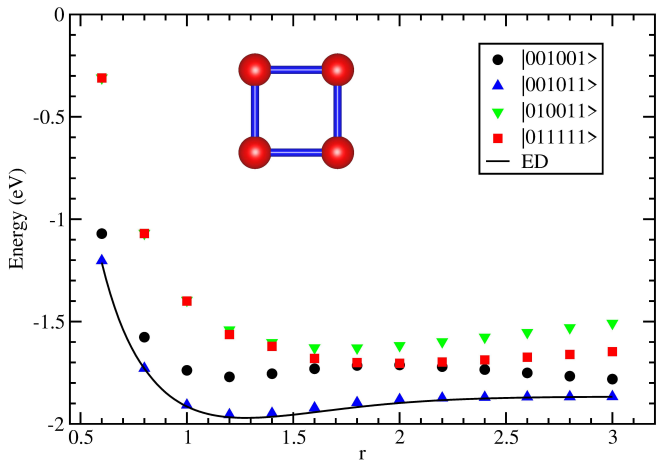}
\caption{\label{h4cluster}(a) The ground state energy of $\mathrm{H}_4$ square as a function of H-H distance, calculated by VQE in 4 subspaces on a 6-qubit statevector simulator, as well as by ED. The inset gives the geometric configuration of the H$_4$ square. }
\end{figure}

Each fermionic operator $U_i$ can be mapped onto a sum of tensor products of Pauli matrices~\cite{Romero2019}. While we have shown that all the terms in the sum can be exactly grouped into a single term in the example of $\mathrm{H_2}$, it is not achievable mathematically in general cases. On the other hand, a full treatment of all the subterms requires a significantly elongated quantum circuit. For instance, a unitarized double excitation operator transforms into 8 subterms in qubit representation~\cite{Romero2019}, which requires a quantum circuit 8 times as long as that for a single term. On a noisy device or simulator where the noise grows cumulatively with the circuit length, it is likely that the extra noise associated with the elongated circuit outweighs the extra accuracy it gains from the extended variational degrees of freedom. Thus, it is helpful to compare the performance of the variational ansatz containing all the subterms for each excitation operator with the one that only contains a single term. 

We performed calculations on $\mathrm{H}_4$ molecule in the square configuration for a wide range of nearest-neighboring H-H distance $r$. The Hamiltonian was constructed based on spin-orbital basis obtained from restricted HF calculations. To ensure that the HF calculation converges to states that are smooth with continuously varying H-H distances, the converged one-particle density matrix at one $r$ was used as the starting point for the next $r$. The 8 spin-orbitals for $\mathrm{H}_4$ were mapped onto a 6-qubit system with parity encoding and subsequent two-qubit reduction. The relevant Hilbert space for $\mathrm{H}_4$ with zero net magnetization has a dimension of $\binom{4}{2}\cdot\binom{4}{2}=36$. Using the graph clustering algorithm, this space can be further separated into 4 subspaces, with dimensions of 8, 8, 10 and 10, respectively. The same partitioning of the Hilbert space can also be obtained for the Hamiltonian constructed based on symmetrized superposition of natural atomic orbitals (i.e., without the self-consistent HF calculations), showing the partitioning comes from the intrinsic point-group symmetry of the molecule, which is $D_{4h}$ in this case. Since our method depends solely on the Hamiltonian matrix, a detailed group theoretical study is not necessary to identify the partitioning of the Hilbert space. 

Fig.~\ref{h4cluster} shows the VQE calculations in which each excitation operator was represented by the first term based on the lexicographical order. For example, among the 8 Pauli strings resulting from the double excitation operator $a^\dagger_8a^\dagger_4a_1a_5$: $\frac{i}{8}(
Y_6X_5X_4X_3X_2X_1+
X_6X_5X_4Y_3X_2X_1+
Y_6X_5Y_4Y_3X_2X_1-
X_6X_5Y_4X_3X_2X_1+
Y_6X_5X_4Y_3X_2Y_1-
X_6X_5X_4X_3X_2Y_1-
Y_6X_5Y_4X_3X_2Y_1-
X_6X_5Y_4Y_3X_2Y_1)$, only the first term $iY_6X_5X_4X_3X_2X_1$ was included in the variational ansatz. VQE calculations were performed in all four subspaces separately, using the noiseless statevector simulator implemented in QISKiT~\cite{Note1}. The statevector simulator uses matrices, rather than Pauli gates, to represent the qubit operators. Thus, it does not involve any noises associated with gate infidelity or imperfect averaging. In each subspace, a reasonable choice of the initial state is the basis state with the lowest diagonal element of the Hamiltonian matrix. An optimized energy close to the ground-state energy calculated by ED can be obtained in the subspace containing $\lvert001011\rangle$ for all H-H distances, clear demonstrating that the ansatz with single-term representation for excitation operators is sufficient to give satisfactory accuracy. We also verified that the final answer is not sensitive to which term was selected. In the following, only single-term operators will be considered unless otherwise noted.

\begin{figure*}
\includegraphics[width=\linewidth]{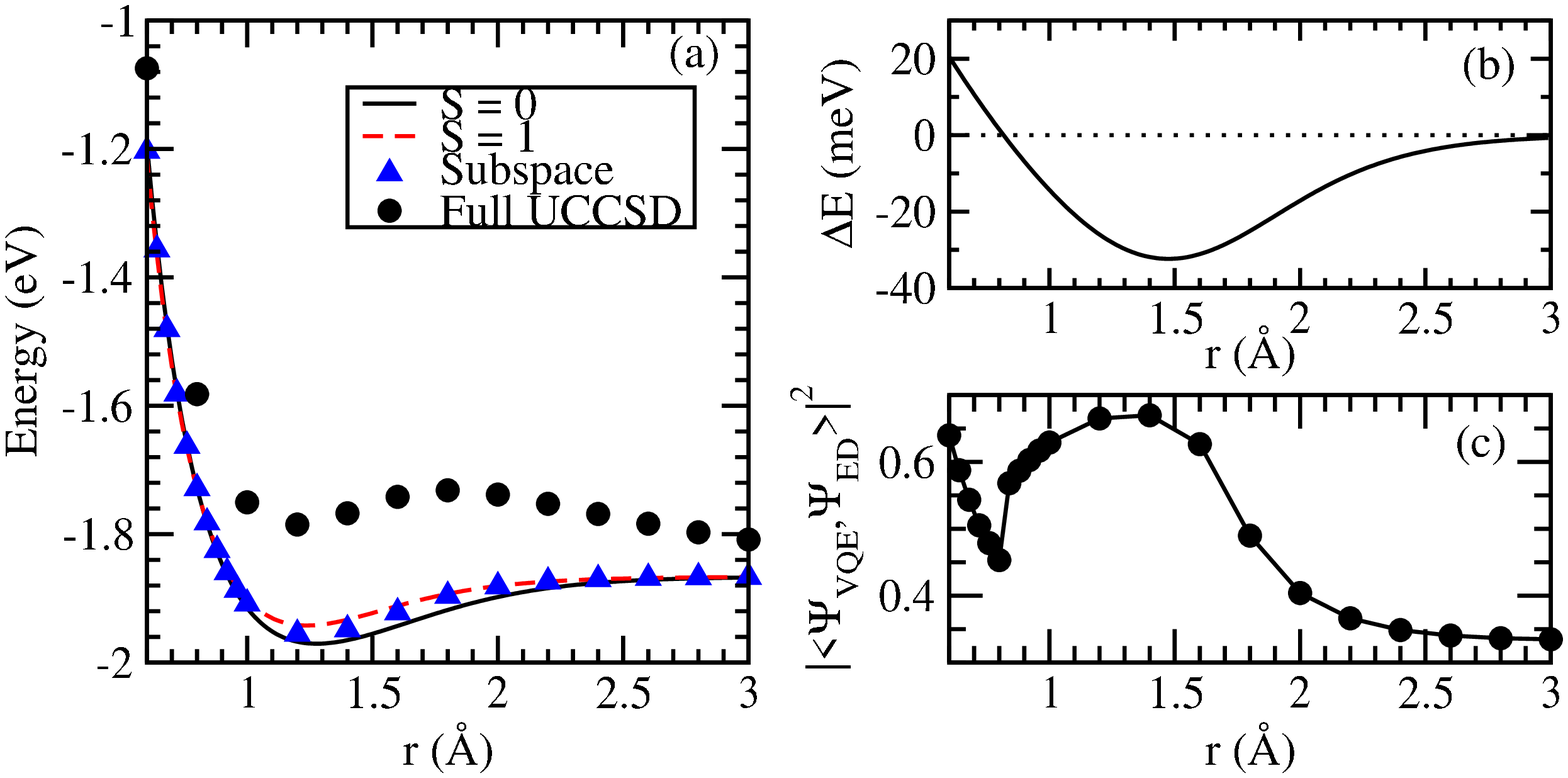}
\caption{\label{spin} (a) Energy of the two lowest-energy eigenstates as a function of $r$ from ED, with $S=0$ (solid black line) and $S=1$ (dashed red line). Also shown are VQE energies calculated in a subspace with selected operators using $\lvert001011\rangle$ as $\lvert\psi_0\rangle$, and calculated in the full Hilbert space with full UCCSD operators using the restricted HF state $\lvert001001\rangle$ as $\lvert\psi_0\rangle$. The 
full UCCSD fails to reach a satisfying estimation of the groun-state energy as the initial state is orthogonal to the ground state. (b) The energy difference between the $S=0$ state and the $S=1$ state. (c) The overlap of the optimized state in VQE performed in the subspace with the ground-state from ED. Near degeneracy between states with different $S^2$ quantum numbers leads to the observed small overlap.}
\end{figure*}

A closer inspection of the ED solution can reveal that there is an energy level crossing between two states with different total angular momentum $S=1$ and $S=0$. In Fig.~\ref{spin} (a), we plot the energy of the two states as a function of the H-H distance. The energy difference $\Delta E=E(S=0)-E(S=1)$ is shown in Fig.~\ref{spin} (b). At bond lengths $r<r_c=0.8$~\AA, the $S=1$ state has the lowest energy. The level crossing occurs at $r_c=0.8$~\AA, where the $S=0$ state becomes more stable. In our VQE calculations, only $E$ is minimized without conserving the $S^2$ quantum numbers. Therefore, due to near degeneracies of $S=0$ and $S=1$ at low energies, the energy optimized VQE wavefunction is spin contaminated.  Since both these two eigenstates are located in the same subspace in which the VQE calculations were performed, and the third eigenstate in this subspace with $S=2$ is distantly separated from the first two states in the vicinity of the equilibrium bond length, the optimized VQE state (blue triangles) is essentially a superposition of the $S=0$ state and the $S=1$ state in this range. In Fig.~\ref{spin} (c), we show the overlap between the VQE state and the ED ground state defined as $\lvert\langle\Psi_\mathrm{VQE}|\Psi_\mathrm{ED}\rangle\rvert^2$. Initially, the overlap drops as the energy difference between the $S=0$ and $S=1$ states decreases. The smallest overlap occurs at $r=0.8$~\AA, which reflects the level crossing when the $S=0$ and $S=1$ states become degenerate. Then, the overlap increases as the $S=0$ and $S=1$ states are separated again, and reaches a maximum at $r=1.4$~\AA, which also coincides with the largest energy difference between the $S=0$ and $S=1$ states. The separation between the two eigenstates shrinks when $r$ further increases. For large $r$ approaching the atomic limit, the energy of the $S=2$ state decreases and eventually becomes degenerate with the first two eigenstates. Consequently, the VQE state becomes a superposition of the three states when approaching this limit: the overlap of the VQE optimized state and the $S=0$ eigenstate decreases to as low as 0.34 at $r=3.0$~\AA. Since our variational ansatz is designed to minimize the energy, it is acceptable that the optimized VQE state is spin contaminated. However, if one wants to preserve the spin quantum numbers, further constraints can be applied in the selection of excitation operators ($T$) to enforce $\left[T,S^2\right]=0$~\cite{Scuseria1991}.  

It is also worth noting that the VQE leading to the lowest-energy solution was performed in a subspace orthogonal to the restricted HF state $\lvert001001\rangle$, using $\lvert001011\rangle$ as the initial state. In fact, the qubit state $\lvert001011\rangle$ can be transformed back to the Fock state $\lvert00110101\rangle$, in which the spin-up electrons (the right half) and spin-down electrons (the left half) do not occupy the same spatial orbitals; while in the restricted HF state $\lvert00110011\rangle$ ($\lvert001001\rangle$ in qubit representation), each molecular orbital is doubly occupied by a spin-up and a spin-down electrons. Without partitioning the Hilbert space, the natural choice is to run VQE with the full set of UCCSD operators using the restricted HF state as the initial state. We showed the results of such calculations on the statevector simulator in Fig.~\ref{spin} (a) as the black circles, where one can see that VQE failed to reach a satisfying estimation of the ground-state energy. This shows that the restricted HF state is not always a good choice as the initial state in VQE calculations, and the partitioning of Hilbert space as performed in the current work can help identify a suitable alternative choice.   

\begin{figure}
\includegraphics[width=\linewidth]{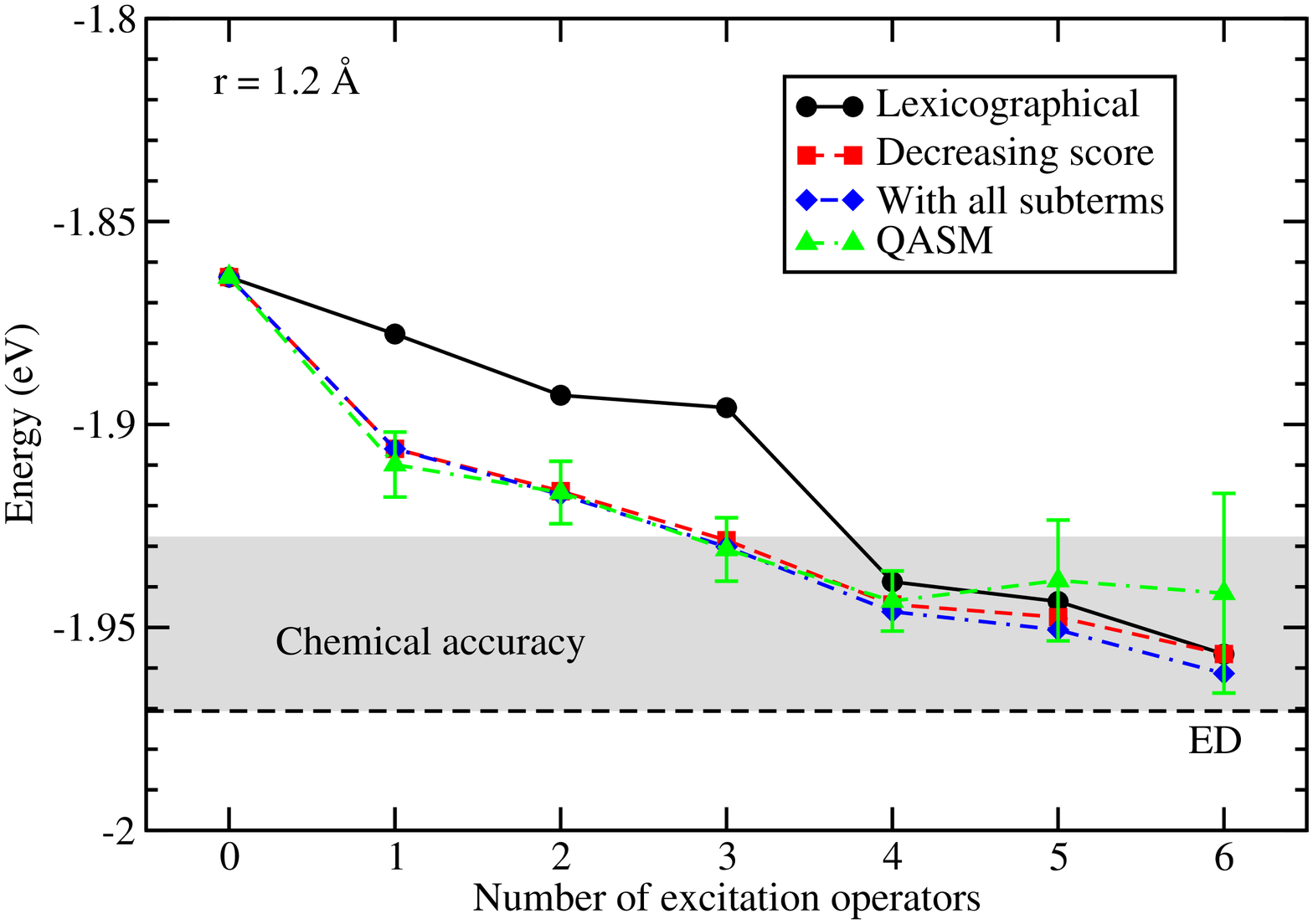} 
\caption{\label{oporder} Energy of $\mathrm{H}_4$-square at at the equilibrium separation $r=1.2$~\AA~as a function of the number of excitation operators, calculated by VQE on statevector and QASM simulators. For statevector simulations, the operators are added according to the lexicographical order or according to decreasing order of the score, respectively. For black circles and red squares, only the first term for each excitation operator is included. The blue diamonds show the statevector simulations that include all subterms for each excitation operator for comparison. The noisy QASM simulation (green triangles) follows the decreasing order of the score and uses the single-term representation for the excitation operators. The shaded gray area shows the ``chemical accuracy" region.}
\end{figure}

The order in which the operators are included in the ansatz also matters, since it is preferable to add operators that create relatively high energy reductions first when the noise is still under control. A related idea of building an effective variational ansatz was illustrated in the ADAPT-VQE algorithm~\cite{Grimsley2019}. Here, we design the following process based on the second-order perturbation theory to efficiently rank the operators. For each $U_i$ connecting $\lvert\psi_0\rangle$ to $\lvert\psi_i\rangle$, we denote the Hamiltonian matrix elements $H_{00}=\epsilon_0$, $H_{ii}=\epsilon_i$ and  $H_{0i}=H_{i0}=\epsilon_{0i}$. Then, we define a ``score" $s_i=\min(\lvert\epsilon_{0i}\rvert,\epsilon_{0i}^2/\lvert\epsilon_{0}-\epsilon_i\rvert)$. The operators will be ranked in the descending order of $s_i$. In Fig.~\ref{oporder}, we show the VQE results by adding excitation operators to the variational ansatz, one at a time according to a lexicographical order as well as decreasing order of $s_i$. The H-H distance was kept at 1.2~\AA, which is the equilibrium distance as seen in Fig.~\ref{h4cluster}. $\lvert001011\rangle$ was set as the initial state. It can be clearly seen that by adding the operators with large $s_i$ first, one can achieve a faster drop of the variational energy at the beginning. Since the noise-free statevector simulator was used, the two different sequences eventually lead to the same energy when all operators are included. The final energy is 0.014 eV above the ED result, which is well within the chemical accuracy of 1 kcal per mole, or 0.043 eV per molecule. Also shown in Fig.~\ref{oporder} are the VQE results by including all subterms for each excitation operator, following the decreasing order of $s_i$. When only one operator was included, the results with a single term or all subterms are exactly the same. On the other hand, when additional operators were added, the VQE energy with all subterms included is slightly lower than that with only a single term. This is expected since more variational degrees of freedom are allowed with the additional Pauli terms. However, the improvement is limited. With all 6 variational operators added, the energy difference between all terms and a single term is only 4.7 meV. Calculations on noisy QASM simulators are also shown in green triangles in Fig.~\ref{oporder}. The error bar was determined based on 10 independent calculations. When 4 or less number of excitation operators were included in the variational ansatz, results consistent with statevector simulators can be achieved on the noisy simulator. However, when the number of operators exceeds 4, the error bar significantly increases, and no further improvement of the energy can be obtained.

\begin{figure*}
\includegraphics[width=\linewidth]{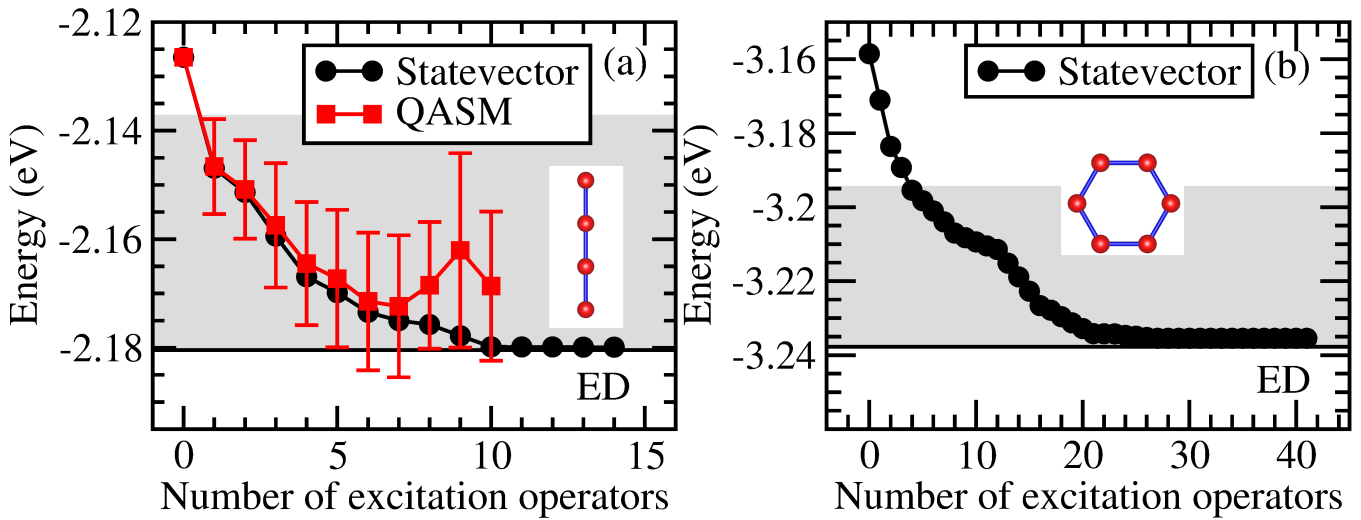} 
\caption{\label{chainhex} (a) Energy of $\mathrm{H}_4$-chain at at the equilibrium separation $r=0.88$~\AA~as a function of the number of excitation operators, calculated by VQE on statevector and QASM simulators. (b) Energy of $\mathrm{H}_6$-hexagon at at the equilibrium separation $r=0.99$~\AA~as a function of the number of excitation operators, calculated by VQE on the statevector simulator. Single-term representation is used for the excitation operators, and the operators are added according to decreasing order of the score. The shade areas show the regions with chemical accuracy. The insets in (a) and (b) show the geometric configurations for H$_4$-chain and H$_6$-hex, respectively.}
\end{figure*}

Two other geometric configurations were studied: an equidistant linear $\mathrm{H}_4$ chain and hexagonal $\mathrm{H}_6$. For the $\mathrm{H}_4$ chain, partitioning of the Hilbert space results in two relevant subspaces with the dimension of 16 and 20, respectively. The ground state can be approached by performing VQE in the 20-dimensional subspace using the HF state as the starting point. A total number of 14 excitation operators were selected and ranked according to the score parameter introduced in the above. As shown in Fig.~\ref{chainhex} (a), on a statevector simulator, 10 excitation operators added in the decreasing order of the score can reach the ED ground-state energy within 1 meV. The remaining 4 excitation operators essentially have no effect on the optimized VQE energy. Therefore, only the first 10 operators were considered in QASM simulations. Again, 10 independent runs were performed for statistical analysis. Similar to the case of $\mathrm{H}_4$-square, only the first 7 excitation operators resulted in reduction of the VQE energy, while further addition of excitation operators did not help because the accumulative noise became too big. Nevertheless, QASM simulations can still reach the lowest energy only 10 meV above the ED result, which is well within the chemical accuracy.  

The hexagonal $\mathrm{H}_6$ can be mapped onto a 10-qubit system with parity encoding and subsequent application of 2-qubit reduction enabled by the Z$_2$ symmetry. The relevant subspace in the qubit system has a dimension of $\binom{6}{3}\cdot\binom{6}{3}=400$ for 6 total electrons and zero net magnetization. This subspace which can be further partitioned into 4 smaller subspaces with dimensions of 96, 96, 104 and 104, respectively. Similar to the $\mathrm{H}_4$-chain case, we identified that the subspace containing the restricted HF state is where the ground state is located. 41 operators up to double excitation can be selected in this subspace. In Fig.~\ref{chainhex} (b), we plot the change of the VQE energy with successive addition of these excitation operators following the decreasing order of the score. One can see that in this case, the score parameter does not fully describe the ``importance" of the operator, since operators 13-16 generate larger energy reductions than the operators immediately preceding them. This is not surprising because underlying perturbation theory can fail when the amplitudes of excitation operators become large. Nevertheless, our scheme still identifies most operators that cause significant energy drop with little extra computational cost. In fact, nearly half the the operators that essentially have no effects on energy reduction were found and put to the end of the list. Alternatively, the ADAPT-VQE algorithm~\cite{Grimsley2019} uses iteratively evaluated gradients of the energy with respect to the operator amplitudes to rank the operators. Since a separate optimization is required in each iteration, this treatment, while being more accurate, is considerably more expensive computationally. 

\begin{figure}
\includegraphics[width=\linewidth]{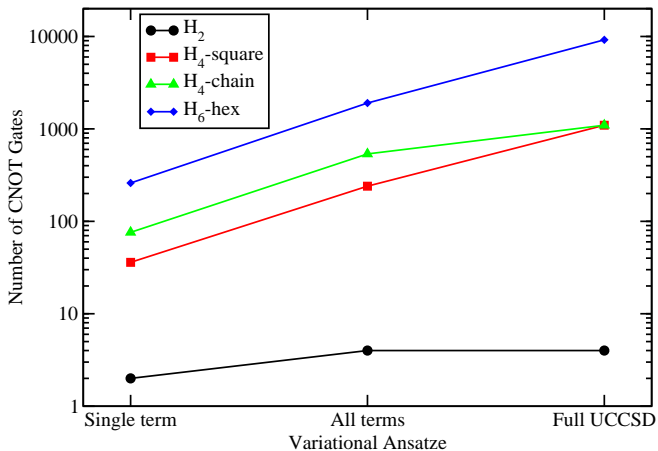}
\caption{\label{circuit} The number of CNOT gates in the quantum circuit to prepare the variational states for ans\"atze with single term for each selected operator in a subspace, all terms for each selected operator in a subspace, and full UCCSD in the nonpartitioned Hilbert space.}
\end{figure}

The CNOT gate is an essential component in gate-based quantum computers; and the number of CNOT gates is a good indicator of the depth of the quantum circuit. In Fig.~\ref{circuit}, we plot the number of CNOT gates in VQE with three different variational ans\"atze: single-term representation in subspace, all-subterm representation in subspace, and the full USSCD. For the hexagonal 
H$_6$, the full UCCSD circuit contains 9,600 CNOT gates, which is out of the working range of the current NISQ. On the other hand, implementing the ansatz with the single-term representation for all operators in the relevant subspace only requires 260 CNOT gates, a reduction by a factor of 35. Since no quantum speedup can be gained on a classical simulator, it is still time-consuming to simulate H$_6$ on the noisy QASM simulator, even with the significant simplification of the circuit: it takes a few hours to prepare the parametrized variational state and measure its expectation value with 1024 shots. For this reason, we chose not to include the QASM calculations on H$_6$ without hurting the main conclusions.

\section{Conclusion}
We introduce a graph clustering algorithm to partition the Hilbert space into subspaces that is made possible by the intrinsic point group symmetry of the molecular systems. This step significantly reduces the number of variational operators since VQE ans\"atze can be confined to act within a particular subspace. Besides, it helps to obtain excitation energies, as shown for the case of $\mathrm{H}_2$), or identify the correct initial state, as demonstrated for $\mathrm{H}_4$. Each excitation operator in UCCSD can be transformed into multiple Pauli terms, requiring a lengthy circuit to represent. We demonstrate with various examples that a single-term representation of excitation operators can reach required accuracy, while dramatically shortening the quantum circuit. VQE calculations on noiseless statevector quantum simulators achieve energies within a few meVs of those obtained with the full USSCD ansatz for $\mathrm{H}_4$ square, $\mathrm{H}_4$ chain and $\mathrm{H}_6$ hexagon molecules. A ``score" parameter was introduced at little extra computational cost, which allows us to rank the excitation operators so that the operators causing larger energy reduction can be applied first. Using $\mathrm{H}_4$-square and $\mathrm{H}_4$-chain as examples, we demonstrate on noisy quantum simulators that only the first few variational operators identified with this strategy are effective in reducing the energy to within the chemical accuracy. 

\begin{acknowledgments}
This work was supported by the U.S. Department of Energy (DOE), Office of Science, Basic Energy Sciences, Materials Science and Engineering Division. The research was performed at the Ames Laboratory, which is operated for the U.S. DOE by Iowa State University under Contract No. DE-AC02-07CH11358.
\end{acknowledgments}

\appendix*
\section{Pseudocode for the graph clustering algorithm}
\begin{itemize}
    \item[] {\bf for} Node\_$i$ in Graph
    \subitem {\bf for} Node\_$j$ in Graph
    \subsubitem ~~~{\bf if} Distance (Node\_$i$, Node\_$j$) $> \epsilon$ {\bf continue}
    \subsubitem ~~~{\bf if} Node\_$i$.cluster == {\bf None} \& Node\_$j$.cluster == {\bf None} 
    \subsubitem ~~~~~~~~~Node\_$i$.cluster = Node\_$j$.cluster = {\bf new} cluster
    \subsubitem ~~~{\bf else if} Node\_$i$.cluster == {\bf None}
    \subsubitem ~~~~~~~~~Node\_$i$.cluster = Node\_$j$.cluster
    \subsubitem ~~~{\bf else if} Node\_$j$.cluster == {\bf None}
    \subsubitem ~~~~~~~~~Node\_$j$.cluster = Node\_$i$.cluster
    \subsubitem ~~~{\bf else if} Node\_$i$.cluster != Node\_$j$.cluster
    \subsubitem ~~~~~~~~~Merge(Node\_$i$.cluster, Node\_$j$.cluster)
    \subsubitem ~~~{\bf end}
    \subitem {\bf end}
    \item[] {\bf end}
\end{itemize}

\bibliography{main}

\begin{thebibliography}{31}%
\makeatletter
\providecommand \@ifxundefined [1]{%
 \@ifx{#1\undefined}
}%
\providecommand \@ifnum [1]{%
 \ifnum #1\expandafter \@firstoftwo
 \else \expandafter \@secondoftwo
 \fi
}%
\providecommand \@ifx [1]{%
 \ifx #1\expandafter \@firstoftwo
 \else \expandafter \@secondoftwo
 \fi
}%
\providecommand \natexlab [1]{#1}%
\providecommand \enquote  [1]{``#1''}%
\providecommand \bibnamefont  [1]{#1}%
\providecommand \bibfnamefont [1]{#1}%
\providecommand \citenamefont [1]{#1}%
\providecommand \href@noop [0]{\@secondoftwo}%
\providecommand \href [0]{\begingroup \@sanitize@url \@href}%
\providecommand \@href[1]{\@@startlink{#1}\@@href}%
\providecommand \@@href[1]{\endgroup#1\@@endlink}%
\providecommand \@sanitize@url [0]{\catcode `\\12\catcode `\$12\catcode
  `\&12\catcode `\#12\catcode `\^12\catcode `\_12\catcode `\%12\relax}%
\providecommand \@@startlink[1]{}%
\providecommand \@@endlink[0]{}%
\providecommand \url  [0]{\begingroup\@sanitize@url \@url }%
\providecommand \@url [1]{\endgroup\@href {#1}{\urlprefix }}%
\providecommand \urlprefix  [0]{URL }%
\providecommand \Eprint [0]{\href }%
\providecommand \doibase [0]{https://doi.org/}%
\providecommand \selectlanguage [0]{\@gobble}%
\providecommand \bibinfo  [0]{\@secondoftwo}%
\providecommand \bibfield  [0]{\@secondoftwo}%
\providecommand \translation [1]{[#1]}%
\providecommand \BibitemOpen [0]{}%
\providecommand \bibitemStop [0]{}%
\providecommand \bibitemNoStop [0]{.\EOS\space}%
\providecommand \EOS [0]{\spacefactor3000\relax}%
\providecommand \BibitemShut  [1]{\csname bibitem#1\endcsname}%
\let\auto@bib@innerbib\@empty
\bibitem [{\citenamefont {Preskill}(2018)}]{Preskill2018}%
  \BibitemOpen
  \bibfield  {author} {\bibinfo {author} {\bibfnamefont {J.}~\bibnamefont
  {Preskill}},\ }\bibfield  {title} {\bibinfo {title} {Quantum {Computing} in
  the {NISQ} era and beyond},\ }\href@noop {} {\bibfield  {journal} {\bibinfo
  {journal} {Quantum}\ }\textbf {\bibinfo {volume} {2}},\ \bibinfo {pages} {79}
  (\bibinfo {year} {2018})}\BibitemShut {NoStop}%
\bibitem [{\citenamefont {Peruzzo}\ \emph {et~al.}(2014)\citenamefont
  {Peruzzo}, \citenamefont {McClean}, \citenamefont {Shadbolt}, \citenamefont
  {Yung}, \citenamefont {Zhou}, \citenamefont {Love}, \citenamefont
  {Aspuru-Guzik},\ and\ \citenamefont {O'Brien}}]{Peruzzo2014}%
  \BibitemOpen
  \bibfield  {author} {\bibinfo {author} {\bibfnamefont {A.}~\bibnamefont
  {Peruzzo}}, \bibinfo {author} {\bibfnamefont {J.}~\bibnamefont {McClean}},
  \bibinfo {author} {\bibfnamefont {P.}~\bibnamefont {Shadbolt}}, \bibinfo
  {author} {\bibfnamefont {M.~H.}\ \bibnamefont {Yung}}, \bibinfo {author}
  {\bibfnamefont {X.~Q.}\ \bibnamefont {Zhou}}, \bibinfo {author}
  {\bibfnamefont {P.~J.}\ \bibnamefont {Love}}, \bibinfo {author}
  {\bibfnamefont {A.}~\bibnamefont {Aspuru-Guzik}},\ and\ \bibinfo {author}
  {\bibfnamefont {J.~L.}\ \bibnamefont {O'Brien}},\ }\bibfield  {title}
  {\bibinfo {title} {{A variational eigenvalue solver on a photonic quantum
  processor}},\ }\bibfield  {journal} {\bibinfo  {journal} {Nature
  Communications}\ }\textbf {\bibinfo {volume} {5}},\ \href
  {https://doi.org/10.1038/ncomms5213} {10.1038/ncomms5213} (\bibinfo {year}
  {2014})\BibitemShut {NoStop}%
\bibitem [{\citenamefont {McArdle}\ \emph {et~al.}(2020)\citenamefont
  {McArdle}, \citenamefont {Endo}, \citenamefont {Aspuru-Guzik}, \citenamefont
  {Benjamin},\ and\ \citenamefont {Yuan}}]{McArdle2020}%
  \BibitemOpen
  \bibfield  {author} {\bibinfo {author} {\bibfnamefont {S.}~\bibnamefont
  {McArdle}}, \bibinfo {author} {\bibfnamefont {S.}~\bibnamefont {Endo}},
  \bibinfo {author} {\bibfnamefont {A.}~\bibnamefont {Aspuru-Guzik}}, \bibinfo
  {author} {\bibfnamefont {S.~C.}\ \bibnamefont {Benjamin}},\ and\ \bibinfo
  {author} {\bibfnamefont {X.}~\bibnamefont {Yuan}},\ }\bibfield  {title}
  {\bibinfo {title} {{Quantum computational chemistry}},\ }\href
  {https://doi.org/10.1103/revmodphys.92.015003} {\bibfield  {journal}
  {\bibinfo  {journal} {Reviews of Modern Physics}\ }\textbf {\bibinfo {volume}
  {92}},\ \bibinfo {pages} {1} (\bibinfo {year} {2020})},\ \Eprint
  {https://arxiv.org/abs/1808.10402} {arXiv:1808.10402} \BibitemShut {NoStop}%
\bibitem [{\citenamefont {Feynman}(1982)}]{Feynman1982}%
  \BibitemOpen
  \bibfield  {author} {\bibinfo {author} {\bibfnamefont {R.~P.}\ \bibnamefont
  {Feynman}},\ }\bibfield  {title} {\bibinfo {title} {{Simulating physics with
  computers}},\ }\href {https://doi.org/10.1007/BF02650179} {\bibfield
  {journal} {\bibinfo  {journal} {International Journal of Theoretical
  Physics}\ }\textbf {\bibinfo {volume} {21}},\ \bibinfo {pages} {467}
  (\bibinfo {year} {1982})}\BibitemShut {NoStop}%
\bibitem [{\citenamefont {Wecker}\ \emph {et~al.}(2015)\citenamefont {Wecker},
  \citenamefont {Hastings},\ and\ \citenamefont {Troyer}}]{Wecker2015}%
  \BibitemOpen
  \bibfield  {author} {\bibinfo {author} {\bibfnamefont {D.}~\bibnamefont
  {Wecker}}, \bibinfo {author} {\bibfnamefont {M.~B.}\ \bibnamefont
  {Hastings}},\ and\ \bibinfo {author} {\bibfnamefont {M.}~\bibnamefont
  {Troyer}},\ }\bibfield  {title} {\bibinfo {title} {Progress towards practical
  quantum variational algorithms},\ }\href
  {https://doi.org/10.1103/PhysRevA.92.042303} {\bibfield  {journal} {\bibinfo
  {journal} {Phys. Rev. A}\ }\textbf {\bibinfo {volume} {92}},\ \bibinfo
  {pages} {042303} (\bibinfo {year} {2015})}\BibitemShut {NoStop}%
\bibitem [{\citenamefont {McClean}\ \emph {et~al.}(2016)\citenamefont
  {McClean}, \citenamefont {Romero}, \citenamefont {Babbush},\ and\
  \citenamefont {Aspuru-Guzik}}]{McClean2016}%
  \BibitemOpen
  \bibfield  {author} {\bibinfo {author} {\bibfnamefont {J.~R.}\ \bibnamefont
  {McClean}}, \bibinfo {author} {\bibfnamefont {J.}~\bibnamefont {Romero}},
  \bibinfo {author} {\bibfnamefont {R.}~\bibnamefont {Babbush}},\ and\ \bibinfo
  {author} {\bibfnamefont {A.}~\bibnamefont {Aspuru-Guzik}},\ }\bibfield
  {title} {\bibinfo {title} {{The theory of variational hybrid
  quantum-classical algorithms}},\ }\href
  {https://doi.org/10.1088/1367-2630/18/2/023023} {\bibfield  {journal}
  {\bibinfo  {journal} {New Journal of Physics}\ }\textbf {\bibinfo {volume}
  {18}},\ \bibinfo {pages} {023023} (\bibinfo {year} {2016})}\BibitemShut
  {NoStop}%
\bibitem [{\citenamefont {O'Malley}\ \emph {et~al.}(2016)\citenamefont
  {O'Malley}, \citenamefont {Babbush}, \citenamefont {Kivlichan}, \citenamefont
  {Romero}, \citenamefont {McClean}, \citenamefont {Barends}, \citenamefont
  {Kelly}, \citenamefont {Roushan}, \citenamefont {Tranter}, \citenamefont
  {Ding}, \citenamefont {Campbell}, \citenamefont {Chen}, \citenamefont {Chen},
  \citenamefont {Chiaro}, \citenamefont {Dunsworth}, \citenamefont {Fowler},
  \citenamefont {Jeffrey}, \citenamefont {Lucero}, \citenamefont {Megrant},
  \citenamefont {Mutus}, \citenamefont {Neeley}, \citenamefont {Neill},
  \citenamefont {Quintana}, \citenamefont {Sank}, \citenamefont {Vainsencher},
  \citenamefont {Wenner}, \citenamefont {White}, \citenamefont {Coveney},
  \citenamefont {Love}, \citenamefont {Neven}, \citenamefont {Aspuru-Guzik},\
  and\ \citenamefont {Martinis}}]{OMalley2016}%
  \BibitemOpen
  \bibfield  {author} {\bibinfo {author} {\bibfnamefont {P.~J.}\ \bibnamefont
  {O'Malley}}, \bibinfo {author} {\bibfnamefont {R.}~\bibnamefont {Babbush}},
  \bibinfo {author} {\bibfnamefont {I.~D.}\ \bibnamefont {Kivlichan}}, \bibinfo
  {author} {\bibfnamefont {J.}~\bibnamefont {Romero}}, \bibinfo {author}
  {\bibfnamefont {J.~R.}\ \bibnamefont {McClean}}, \bibinfo {author}
  {\bibfnamefont {R.}~\bibnamefont {Barends}}, \bibinfo {author} {\bibfnamefont
  {J.}~\bibnamefont {Kelly}}, \bibinfo {author} {\bibfnamefont
  {P.}~\bibnamefont {Roushan}}, \bibinfo {author} {\bibfnamefont
  {A.}~\bibnamefont {Tranter}}, \bibinfo {author} {\bibfnamefont
  {N.}~\bibnamefont {Ding}}, \bibinfo {author} {\bibfnamefont {B.}~\bibnamefont
  {Campbell}}, \bibinfo {author} {\bibfnamefont {Y.}~\bibnamefont {Chen}},
  \bibinfo {author} {\bibfnamefont {Z.}~\bibnamefont {Chen}}, \bibinfo {author}
  {\bibfnamefont {B.}~\bibnamefont {Chiaro}}, \bibinfo {author} {\bibfnamefont
  {A.}~\bibnamefont {Dunsworth}}, \bibinfo {author} {\bibfnamefont {A.~G.}\
  \bibnamefont {Fowler}}, \bibinfo {author} {\bibfnamefont {E.}~\bibnamefont
  {Jeffrey}}, \bibinfo {author} {\bibfnamefont {E.}~\bibnamefont {Lucero}},
  \bibinfo {author} {\bibfnamefont {A.}~\bibnamefont {Megrant}}, \bibinfo
  {author} {\bibfnamefont {J.~Y.}\ \bibnamefont {Mutus}}, \bibinfo {author}
  {\bibfnamefont {M.}~\bibnamefont {Neeley}}, \bibinfo {author} {\bibfnamefont
  {C.}~\bibnamefont {Neill}}, \bibinfo {author} {\bibfnamefont
  {C.}~\bibnamefont {Quintana}}, \bibinfo {author} {\bibfnamefont
  {D.}~\bibnamefont {Sank}}, \bibinfo {author} {\bibfnamefont {A.}~\bibnamefont
  {Vainsencher}}, \bibinfo {author} {\bibfnamefont {J.}~\bibnamefont {Wenner}},
  \bibinfo {author} {\bibfnamefont {T.~C.}\ \bibnamefont {White}}, \bibinfo
  {author} {\bibfnamefont {P.~V.}\ \bibnamefont {Coveney}}, \bibinfo {author}
  {\bibfnamefont {P.~J.}\ \bibnamefont {Love}}, \bibinfo {author}
  {\bibfnamefont {H.}~\bibnamefont {Neven}}, \bibinfo {author} {\bibfnamefont
  {A.}~\bibnamefont {Aspuru-Guzik}},\ and\ \bibinfo {author} {\bibfnamefont
  {J.~M.}\ \bibnamefont {Martinis}},\ }\bibfield  {title} {\bibinfo {title}
  {{Scalable quantum simulation of molecular energies}},\ }\href
  {https://doi.org/10.1103/PhysRevX.6.031007} {\bibfield  {journal} {\bibinfo
  {journal} {Physical Review X}\ }\textbf {\bibinfo {volume} {6}},\ \bibinfo
  {pages} {1} (\bibinfo {year} {2016})},\ \Eprint
  {https://arxiv.org/abs/1512.06860} {arXiv:1512.06860} \BibitemShut {NoStop}%
\bibitem [{\citenamefont {McClean}\ \emph {et~al.}(2017)\citenamefont
  {McClean}, \citenamefont {Kimchi-Schwartz}, \citenamefont {Carter},\ and\
  \citenamefont {de~Jong}}]{McClean2017}%
  \BibitemOpen
  \bibfield  {author} {\bibinfo {author} {\bibfnamefont {J.~R.}\ \bibnamefont
  {McClean}}, \bibinfo {author} {\bibfnamefont {M.~E.}\ \bibnamefont
  {Kimchi-Schwartz}}, \bibinfo {author} {\bibfnamefont {J.}~\bibnamefont
  {Carter}},\ and\ \bibinfo {author} {\bibfnamefont {W.~A.}\ \bibnamefont
  {de~Jong}},\ }\bibfield  {title} {\bibinfo {title} {{Hybrid quantum-classical
  hierarchy for mitigation of decoherence and determination of excited
  states}},\ }\href {https://doi.org/10.1103/PhysRevA.95.042308} {\bibfield
  {journal} {\bibinfo  {journal} {Physical Review A}\ }\textbf {\bibinfo
  {volume} {95}},\ \bibinfo {pages} {42308} (\bibinfo {year}
  {2017})}\BibitemShut {NoStop}%
\bibitem [{\citenamefont {Barkoutsos}\ \emph {et~al.}(2018)\citenamefont
  {Barkoutsos}, \citenamefont {Gonthier}, \citenamefont {Sokolov},
  \citenamefont {Moll}, \citenamefont {Salis}, \citenamefont {Fuhrer},
  \citenamefont {Ganzhorn}, \citenamefont {Egger}, \citenamefont {Troyer},
  \citenamefont {Mezzacapo}, \citenamefont {Filipp},\ and\ \citenamefont
  {Tavernelli}}]{Barkoutsos2018}%
  \BibitemOpen
  \bibfield  {author} {\bibinfo {author} {\bibfnamefont {P.~K.}\ \bibnamefont
  {Barkoutsos}}, \bibinfo {author} {\bibfnamefont {J.~F.}\ \bibnamefont
  {Gonthier}}, \bibinfo {author} {\bibfnamefont {I.}~\bibnamefont {Sokolov}},
  \bibinfo {author} {\bibfnamefont {N.}~\bibnamefont {Moll}}, \bibinfo {author}
  {\bibfnamefont {G.}~\bibnamefont {Salis}}, \bibinfo {author} {\bibfnamefont
  {A.}~\bibnamefont {Fuhrer}}, \bibinfo {author} {\bibfnamefont
  {M.}~\bibnamefont {Ganzhorn}}, \bibinfo {author} {\bibfnamefont {D.~J.}\
  \bibnamefont {Egger}}, \bibinfo {author} {\bibfnamefont {M.}~\bibnamefont
  {Troyer}}, \bibinfo {author} {\bibfnamefont {A.}~\bibnamefont {Mezzacapo}},
  \bibinfo {author} {\bibfnamefont {S.}~\bibnamefont {Filipp}},\ and\ \bibinfo
  {author} {\bibfnamefont {I.}~\bibnamefont {Tavernelli}},\ }\bibfield  {title}
  {\bibinfo {title} {{Quantum algorithms for electronic structure calculations:
  Particle-hole Hamiltonian and optimized wave-function expansions}},\ }\href
  {https://doi.org/10.1103/PhysRevA.98.022322} {\bibfield  {journal} {\bibinfo
  {journal} {Physical Review A}\ }\textbf {\bibinfo {volume} {98}},\ \bibinfo
  {pages} {22322} (\bibinfo {year} {2018})}\BibitemShut {NoStop}%
\bibitem [{\citenamefont {Colless}\ \emph {et~al.}(2018)\citenamefont
  {Colless}, \citenamefont {Ramasesh}, \citenamefont {Dahlen}, \citenamefont
  {Blok}, \citenamefont {Kimchi-Schwartz}, \citenamefont {McClean},
  \citenamefont {Carter}, \citenamefont {de~Jong},\ and\ \citenamefont
  {Siddiqi}}]{Colless2018}%
  \BibitemOpen
  \bibfield  {author} {\bibinfo {author} {\bibfnamefont {J.}~\bibnamefont
  {Colless}}, \bibinfo {author} {\bibfnamefont {V.}~\bibnamefont {Ramasesh}},
  \bibinfo {author} {\bibfnamefont {D.}~\bibnamefont {Dahlen}}, \bibinfo
  {author} {\bibfnamefont {M.}~\bibnamefont {Blok}}, \bibinfo {author}
  {\bibfnamefont {M.}~\bibnamefont {Kimchi-Schwartz}}, \bibinfo {author}
  {\bibfnamefont {J.}~\bibnamefont {McClean}}, \bibinfo {author} {\bibfnamefont
  {J.}~\bibnamefont {Carter}}, \bibinfo {author} {\bibfnamefont
  {W.}~\bibnamefont {de~Jong}},\ and\ \bibinfo {author} {\bibfnamefont
  {I.}~\bibnamefont {Siddiqi}},\ }\bibfield  {title} {\bibinfo {title}
  {{Computation of Molecular Spectra on a Quantum Processor with an
  Error-Resilient Algorithm}},\ }\href
  {https://doi.org/10.1103/PhysRevX.8.011021} {\bibfield  {journal} {\bibinfo
  {journal} {Physical Review X}\ }\textbf {\bibinfo {volume} {8}},\ \bibinfo
  {pages} {11021} (\bibinfo {year} {2018})}\BibitemShut {NoStop}%
\bibitem [{\citenamefont {Romero}\ \emph {et~al.}(2019)\citenamefont {Romero},
  \citenamefont {Babbush}, \citenamefont {McClean}, \citenamefont {Hempel},
  \citenamefont {Love},\ and\ \citenamefont {Aspuru-Guzik}}]{Romero2019}%
  \BibitemOpen
  \bibfield  {author} {\bibinfo {author} {\bibfnamefont {J.}~\bibnamefont
  {Romero}}, \bibinfo {author} {\bibfnamefont {R.}~\bibnamefont {Babbush}},
  \bibinfo {author} {\bibfnamefont {J.~R.}\ \bibnamefont {McClean}}, \bibinfo
  {author} {\bibfnamefont {C.}~\bibnamefont {Hempel}}, \bibinfo {author}
  {\bibfnamefont {P.~J.}\ \bibnamefont {Love}},\ and\ \bibinfo {author}
  {\bibfnamefont {A.}~\bibnamefont {Aspuru-Guzik}},\ }\bibfield  {title}
  {\bibinfo {title} {{Strategies for quantum computing molecular energies using
  the unitary coupled cluster ansatz}},\ }\bibfield  {journal} {\bibinfo
  {journal} {Quantum Science and Technology}\ }\textbf {\bibinfo {volume}
  {4}},\ \href {https://doi.org/10.1088/2058-9565/aad3e4}
  {10.1088/2058-9565/aad3e4} (\bibinfo {year} {2019}),\ \Eprint
  {https://arxiv.org/abs/1701.02691} {arXiv:1701.02691} \BibitemShut {NoStop}%
\bibitem [{\citenamefont {Higgott}\ \emph {et~al.}(2019)\citenamefont
  {Higgott}, \citenamefont {Wang},\ and\ \citenamefont
  {Brierley}}]{Higgott2019}%
  \BibitemOpen
  \bibfield  {author} {\bibinfo {author} {\bibfnamefont {O.}~\bibnamefont
  {Higgott}}, \bibinfo {author} {\bibfnamefont {D.}~\bibnamefont {Wang}},\ and\
  \bibinfo {author} {\bibfnamefont {S.}~\bibnamefont {Brierley}},\ }\bibfield
  {title} {\bibinfo {title} {{Variational Quantum Computation of Excited
  States}},\ }\href {https://doi.org/10.22331/q-2019-07-01-156} {\bibfield
  {journal} {\bibinfo  {journal} {Quantum}\ }\textbf {\bibinfo {volume} {3}},\
  \bibinfo {pages} {156} (\bibinfo {year} {2019})},\ \Eprint
  {https://arxiv.org/abs/1805.08138} {arXiv:1805.08138} \BibitemShut {NoStop}%
\bibitem [{\citenamefont {Lee}\ \emph {et~al.}(2019)\citenamefont {Lee},
  \citenamefont {Huggins}, \citenamefont {Head-Gordon},\ and\ \citenamefont
  {Whaley}}]{Lee2019}%
  \BibitemOpen
  \bibfield  {author} {\bibinfo {author} {\bibfnamefont {J.}~\bibnamefont
  {Lee}}, \bibinfo {author} {\bibfnamefont {W.~J.}\ \bibnamefont {Huggins}},
  \bibinfo {author} {\bibfnamefont {M.}~\bibnamefont {Head-Gordon}},\ and\
  \bibinfo {author} {\bibfnamefont {K.~B.}\ \bibnamefont {Whaley}},\ }\bibfield
   {title} {\bibinfo {title} {{Generalized Unitary Coupled Cluster Wave
  functions for Quantum Computation}},\ }\href
  {https://doi.org/10.1021/acs.jctc.8b01004} {\bibfield  {journal} {\bibinfo
  {journal} {Journal of Chemical Theory and Computation}\ }\textbf {\bibinfo
  {volume} {15}},\ \bibinfo {pages} {311} (\bibinfo {year} {2019})},\ \Eprint
  {https://arxiv.org/abs/1810.02327} {arXiv:1810.02327} \BibitemShut {NoStop}%
\bibitem [{\citenamefont {Grimsley}\ \emph {et~al.}(2019)\citenamefont
  {Grimsley}, \citenamefont {Economou}, \citenamefont {Barnes},\ and\
  \citenamefont {Mayhall}}]{Grimsley2019}%
  \BibitemOpen
  \bibfield  {author} {\bibinfo {author} {\bibfnamefont {H.~R.}\ \bibnamefont
  {Grimsley}}, \bibinfo {author} {\bibfnamefont {S.~E.}\ \bibnamefont
  {Economou}}, \bibinfo {author} {\bibfnamefont {E.}~\bibnamefont {Barnes}},\
  and\ \bibinfo {author} {\bibfnamefont {N.~J.}\ \bibnamefont {Mayhall}},\
  }\bibfield  {title} {\bibinfo {title} {{An adaptive variational algorithm for
  exact molecular simulations on a quantum computer}},\ }\bibfield  {journal}
  {\bibinfo  {journal} {Nature Communications}\ }\textbf {\bibinfo {volume}
  {10}},\ \href {https://doi.org/10.1038/s41467-019-10988-2}
  {10.1038/s41467-019-10988-2} (\bibinfo {year} {2019}),\ \Eprint
  {https://arxiv.org/abs/1812.11173} {arXiv:1812.11173} \BibitemShut {NoStop}%
\bibitem [{\citenamefont {Tang}\ \emph {et~al.}(2019)\citenamefont {Tang},
  \citenamefont {Barnes}, \citenamefont {Grimsley}, \citenamefont {Mayhall},\
  and\ \citenamefont {Economou}}]{Tang2019}%
  \BibitemOpen
  \bibfield  {author} {\bibinfo {author} {\bibfnamefont {H.~L.}\ \bibnamefont
  {Tang}}, \bibinfo {author} {\bibfnamefont {E.}~\bibnamefont {Barnes}},
  \bibinfo {author} {\bibfnamefont {H.~R.}\ \bibnamefont {Grimsley}}, \bibinfo
  {author} {\bibfnamefont {N.~J.}\ \bibnamefont {Mayhall}},\ and\ \bibinfo
  {author} {\bibfnamefont {S.~E.}\ \bibnamefont {Economou}},\ }\bibfield
  {title} {\bibinfo {title} {{qubit-ADAPT-VQE: An adaptive algorithm for
  constructing hardware-efficient ansatze on a quantum processor}},\ }\href
  {http://arxiv.org/abs/1911.10205} {\ ,\ \bibinfo {pages} {1} (\bibinfo {year}
  {2019})},\ \Eprint {https://arxiv.org/abs/1911.10205} {arXiv:1911.10205}
  \BibitemShut {NoStop}%
\bibitem [{\citenamefont {Kandala}\ \emph {et~al.}(2017)\citenamefont
  {Kandala}, \citenamefont {Mezzacapo}, \citenamefont {Temme}, \citenamefont
  {Takita}, \citenamefont {Brink}, \citenamefont {Chow},\ and\ \citenamefont
  {Gambetta}}]{Kandala2017}%
  \BibitemOpen
  \bibfield  {author} {\bibinfo {author} {\bibfnamefont {A.}~\bibnamefont
  {Kandala}}, \bibinfo {author} {\bibfnamefont {A.}~\bibnamefont {Mezzacapo}},
  \bibinfo {author} {\bibfnamefont {K.}~\bibnamefont {Temme}}, \bibinfo
  {author} {\bibfnamefont {M.}~\bibnamefont {Takita}}, \bibinfo {author}
  {\bibfnamefont {M.}~\bibnamefont {Brink}}, \bibinfo {author} {\bibfnamefont
  {J.~M.}\ \bibnamefont {Chow}},\ and\ \bibinfo {author} {\bibfnamefont
  {J.~M.}\ \bibnamefont {Gambetta}},\ }\bibfield  {title} {\bibinfo {title}
  {{Hardware-efficient variational quantum eigensolver for small molecules and
  quantum magnets}},\ }\href {https://doi.org/10.1038/nature23879} {\bibfield
  {journal} {\bibinfo  {journal} {Nature}\ }\textbf {\bibinfo {volume} {549}},\
  \bibinfo {pages} {242} (\bibinfo {year} {2017})}\BibitemShut {NoStop}%
\bibitem [{\citenamefont {Hempel}\ \emph {et~al.}(2018)\citenamefont {Hempel},
  \citenamefont {Maier}, \citenamefont {Romero}, \citenamefont {McClean},
  \citenamefont {Monz}, \citenamefont {Shen}, \citenamefont {Jurcevic},
  \citenamefont {Lanyon}, \citenamefont {Love}, \citenamefont {Babbush},
  \citenamefont {Aspuru-Guzik}, \citenamefont {Blatt},\ and\ \citenamefont
  {Roos}}]{Hempel2018}%
  \BibitemOpen
  \bibfield  {author} {\bibinfo {author} {\bibfnamefont {C.}~\bibnamefont
  {Hempel}}, \bibinfo {author} {\bibfnamefont {C.}~\bibnamefont {Maier}},
  \bibinfo {author} {\bibfnamefont {J.}~\bibnamefont {Romero}}, \bibinfo
  {author} {\bibfnamefont {J.}~\bibnamefont {McClean}}, \bibinfo {author}
  {\bibfnamefont {T.}~\bibnamefont {Monz}}, \bibinfo {author} {\bibfnamefont
  {H.}~\bibnamefont {Shen}}, \bibinfo {author} {\bibfnamefont {P.}~\bibnamefont
  {Jurcevic}}, \bibinfo {author} {\bibfnamefont {B.~P.}\ \bibnamefont
  {Lanyon}}, \bibinfo {author} {\bibfnamefont {P.}~\bibnamefont {Love}},
  \bibinfo {author} {\bibfnamefont {R.}~\bibnamefont {Babbush}}, \bibinfo
  {author} {\bibfnamefont {A.}~\bibnamefont {Aspuru-Guzik}}, \bibinfo {author}
  {\bibfnamefont {R.}~\bibnamefont {Blatt}},\ and\ \bibinfo {author}
  {\bibfnamefont {C.~F.}\ \bibnamefont {Roos}},\ }\bibfield  {title} {\bibinfo
  {title} {{Quantum Chemistry Calculations on a Trapped-Ion Quantum
  Simulator}},\ }\href {https://doi.org/10.1103/PhysRevX.8.031022} {\bibfield
  {journal} {\bibinfo  {journal} {Physical Review X}\ }\textbf {\bibinfo
  {volume} {8}},\ \bibinfo {pages} {31022} (\bibinfo {year} {2018})},\ \Eprint
  {https://arxiv.org/abs/1803.10238} {arXiv:1803.10238} \BibitemShut {NoStop}%
\bibitem [{\citenamefont {Shen}\ \emph {et~al.}(2017)\citenamefont {Shen},
  \citenamefont {Zhang}, \citenamefont {Zhang}, \citenamefont {Zhang},
  \citenamefont {Yung},\ and\ \citenamefont {Kim}}]{Shen2017}%
  \BibitemOpen
  \bibfield  {author} {\bibinfo {author} {\bibfnamefont {Y.}~\bibnamefont
  {Shen}}, \bibinfo {author} {\bibfnamefont {X.}~\bibnamefont {Zhang}},
  \bibinfo {author} {\bibfnamefont {S.}~\bibnamefont {Zhang}}, \bibinfo
  {author} {\bibfnamefont {J.-N.}\ \bibnamefont {Zhang}}, \bibinfo {author}
  {\bibfnamefont {M.-H.}\ \bibnamefont {Yung}},\ and\ \bibinfo {author}
  {\bibfnamefont {K.}~\bibnamefont {Kim}},\ }\bibfield  {title} {\bibinfo
  {title} {{Quantum implementation of the unitary coupled cluster for
  simulating molecular electronic structure}},\ }\href
  {https://doi.org/10.1103/PhysRevA.95.020501} {\bibfield  {journal} {\bibinfo
  {journal} {Physical Review A}\ }\textbf {\bibinfo {volume} {95}},\ \bibinfo
  {pages} {20501} (\bibinfo {year} {2017})}\BibitemShut {NoStop}%
\bibitem [{\citenamefont {Yao}\ \emph {et~al.}(2020)\citenamefont {Yao},
  \citenamefont {Zhang}, \citenamefont {Wang}, \citenamefont {Ho},\ and\
  \citenamefont {Orth}}]{Yao2020}%
  \BibitemOpen
  \bibfield  {author} {\bibinfo {author} {\bibfnamefont {Y.}~\bibnamefont
  {Yao}}, \bibinfo {author} {\bibfnamefont {F.}~\bibnamefont {Zhang}}, \bibinfo
  {author} {\bibfnamefont {C.-Z.}\ \bibnamefont {Wang}}, \bibinfo {author}
  {\bibfnamefont {K.-M.}\ \bibnamefont {Ho}},\ and\ \bibinfo {author}
  {\bibfnamefont {P.~P.}\ \bibnamefont {Orth}},\ }\bibfield  {title} {\bibinfo
  {title} {{Gutzwiller Hybrid Quantum-Classical Computing Approach for
  Correlated Materials}},\ }\href {http://arxiv.org/abs/2003.04211} {\
  (\bibinfo {year} {2020})},\ \Eprint {https://arxiv.org/abs/2003.04211}
  {arXiv:2003.04211} \BibitemShut {NoStop}%
\bibitem [{\citenamefont {Hoffmann}\ and\ \citenamefont
  {Jack}(1988)}]{Hoffmann1988}%
  \BibitemOpen
  \bibfield  {author} {\bibinfo {author} {\bibfnamefont {M.~R.}\ \bibnamefont
  {Hoffmann}}\ and\ \bibinfo {author} {\bibfnamefont {S.}~\bibnamefont
  {Jack}},\ }\bibfield  {title} {\bibinfo {title} {{A unitary
  multiconfigurational coupled- cluster method : Theory and applications and
  applications}},\ }\href@noop {} {\bibfield  {journal} {\bibinfo  {journal} {J
  Chem Phys}\ }\textbf {\bibinfo {volume} {88}},\ \bibinfo {pages} {993}
  (\bibinfo {year} {1988})}\BibitemShut {NoStop}%
\bibitem [{\citenamefont {Bartlett}\ \emph {et~al.}(1989)\citenamefont
  {Bartlett}, \citenamefont {Kucharski},\ and\ \citenamefont
  {Noga}}]{Bartlett1989}%
  \BibitemOpen
  \bibfield  {author} {\bibinfo {author} {\bibfnamefont {R.~J.}\ \bibnamefont
  {Bartlett}}, \bibinfo {author} {\bibfnamefont {S.~A.}\ \bibnamefont
  {Kucharski}},\ and\ \bibinfo {author} {\bibfnamefont {J.}~\bibnamefont
  {Noga}},\ }\bibfield  {title} {\bibinfo {title} {{Alternative Coupled-cluster
  Ansatze II. The Unitary Coupled-cluster Method}},\ }\href@noop {} {\bibfield
  {journal} {\bibinfo  {journal} {Chemical Physics Letters}\ }\textbf {\bibinfo
  {volume} {155}},\ \bibinfo {pages} {133} (\bibinfo {year}
  {1989})}\BibitemShut {NoStop}%
\bibitem [{\citenamefont {Helgaker}\ \emph {et~al.}(2014)\citenamefont
  {Helgaker}, \citenamefont {Jorgensen},\ and\ \citenamefont
  {Olsen}}]{Helgaker2014}%
  \BibitemOpen
  \bibfield  {author} {\bibinfo {author} {\bibfnamefont {T.}~\bibnamefont
  {Helgaker}}, \bibinfo {author} {\bibfnamefont {P.}~\bibnamefont
  {Jorgensen}},\ and\ \bibinfo {author} {\bibfnamefont {J.}~\bibnamefont
  {Olsen}},\ }\href@noop {} {\emph {\bibinfo {title} {{Molecular
  electronic-structure theory}}}}\ (\bibinfo  {publisher} {John Wiley {\&}
  Sons, Inc.},\ \bibinfo {year} {2014})\BibitemShut {NoStop}%
\bibitem [{\citenamefont {Trotter}(1959)}]{Trotter1959}%
  \BibitemOpen
  \bibfield  {author} {\bibinfo {author} {\bibfnamefont {H.~F.}\ \bibnamefont
  {Trotter}},\ }\bibfield  {title} {\bibinfo {title} {{On the Product of
  Semi-Groups of Operators}},\ }\href@noop {} {\bibfield  {journal} {\bibinfo
  {journal} {Proceedings of the American Mathematical Society}\ }\textbf
  {\bibinfo {volume} {10}},\ \bibinfo {pages} {545} (\bibinfo {year}
  {1959})}\BibitemShut {NoStop}%
\bibitem [{\citenamefont {Lanat{\`a}}\ \emph {et~al.}(2015)\citenamefont
  {Lanat{\`a}}, \citenamefont {Yao}, \citenamefont {Wang}, \citenamefont {Ho},\
  and\ \citenamefont {Kotliar}}]{ga_pu}%
  \BibitemOpen
  \bibfield  {author} {\bibinfo {author} {\bibfnamefont {N.}~\bibnamefont
  {Lanat{\`a}}}, \bibinfo {author} {\bibfnamefont {Y.}~\bibnamefont {Yao}},
  \bibinfo {author} {\bibfnamefont {C.-Z.}\ \bibnamefont {Wang}}, \bibinfo
  {author} {\bibfnamefont {K.-M.}\ \bibnamefont {Ho}},\ and\ \bibinfo {author}
  {\bibfnamefont {G.}~\bibnamefont {Kotliar}},\ }\bibfield  {title} {\bibinfo
  {title} {Phase diagram and electronic structure of praseodymium and
  plutonium},\ }\href@noop {} {\bibfield  {journal} {\bibinfo  {journal} {Phys.
  Rev. X}\ }\textbf {\bibinfo {volume} {5}},\ \bibinfo {pages} {011008}
  (\bibinfo {year} {2015})}\BibitemShut {NoStop}%
\bibitem [{\citenamefont {Lanat{\`{a}}}\ \emph {et~al.}(2017)\citenamefont
  {Lanat{\`{a}}}, \citenamefont {Yao}, \citenamefont {Deng}, \citenamefont
  {Dobrosavljevi{\'{c}}},\ and\ \citenamefont {Kotliar}}]{Lanata2017}%
  \BibitemOpen
  \bibfield  {author} {\bibinfo {author} {\bibfnamefont {N.}~\bibnamefont
  {Lanat{\`{a}}}}, \bibinfo {author} {\bibfnamefont {Y.}~\bibnamefont {Yao}},
  \bibinfo {author} {\bibfnamefont {X.}~\bibnamefont {Deng}}, \bibinfo {author}
  {\bibfnamefont {V.}~\bibnamefont {Dobrosavljevi{\'{c}}}},\ and\ \bibinfo
  {author} {\bibfnamefont {G.}~\bibnamefont {Kotliar}},\ }\bibfield  {title}
  {\bibinfo {title} {{Slave Boson Theory of Orbital Differentiation with
  Crystal Field Effects: Application to UO2}},\ }\href
  {https://doi.org/10.1103/PhysRevLett.118.126401} {\bibfield  {journal}
  {\bibinfo  {journal} {Physical Review Letters}\ }\textbf {\bibinfo {volume}
  {118}},\ \bibinfo {pages} {1} (\bibinfo {year} {2017})},\ \Eprint
  {https://arxiv.org/abs/1606.09614} {arXiv:1606.09614} \BibitemShut {NoStop}%
\bibitem [{\citenamefont {Fischer}\ and\ \citenamefont
  {Gunlycke}(2019)}]{Fischer2019}%
  \BibitemOpen
  \bibfield  {author} {\bibinfo {author} {\bibfnamefont {S.~A.}\ \bibnamefont
  {Fischer}}\ and\ \bibinfo {author} {\bibfnamefont {D.}~\bibnamefont
  {Gunlycke}},\ }\bibfield  {title} {\bibinfo {title} {{Symmetry Configuration
  Mapping for Representing Quantum Systems on Quantum Computers}},\ }\href
  {http://arxiv.org/abs/1907.01493} {\  (\bibinfo {year} {2019})},\ \Eprint
  {https://arxiv.org/abs/1907.01493} {arXiv:1907.01493} \BibitemShut {NoStop}%
\bibitem [{Note1()}]{Note1}%
  \BibitemOpen
  \bibinfo {note} {Python api. https://qiskit.org/}\BibitemShut {NoStop}%
\bibitem [{\citenamefont {Hehre}\ \emph {et~al.}(1969)\citenamefont {Hehre},
  \citenamefont {Stewart},\ and\ \citenamefont {Pople}}]{Hehre1969}%
  \BibitemOpen
  \bibfield  {author} {\bibinfo {author} {\bibfnamefont {W.~J.}\ \bibnamefont
  {Hehre}}, \bibinfo {author} {\bibfnamefont {R.~F.}\ \bibnamefont {Stewart}},\
  and\ \bibinfo {author} {\bibfnamefont {J.~A.}\ \bibnamefont {Pople}},\
  }\bibfield  {title} {\bibinfo {title} {{Self-consistent molecular-orbital
  methods. I. Use of gaussian expansions of slater-type atomic orbitals}},\
  }\href {https://doi.org/10.1063/1.1672392} {\bibfield  {journal} {\bibinfo
  {journal} {The Journal of Chemical Physics}\ }\textbf {\bibinfo {volume}
  {51}},\ \bibinfo {pages} {2657} (\bibinfo {year} {1969})}\BibitemShut
  {NoStop}%
\bibitem [{\citenamefont {Sun}\ \emph {et~al.}(2017)\citenamefont {Sun},
  \citenamefont {Berkelbach}, \citenamefont {Blunt}, \citenamefont {Booth},
  \citenamefont {Guo}, \citenamefont {Li}, \citenamefont {Liu}, \citenamefont
  {McClain}, \citenamefont {Sayfutyarova}, \citenamefont {Sharma},
  \citenamefont {Wouters},\ and\ \citenamefont {Chan}}]{PYSCF}%
  \BibitemOpen
  \bibfield  {author} {\bibinfo {author} {\bibfnamefont {Q.}~\bibnamefont
  {Sun}}, \bibinfo {author} {\bibfnamefont {T.~C.}\ \bibnamefont {Berkelbach}},
  \bibinfo {author} {\bibfnamefont {N.~S.}\ \bibnamefont {Blunt}}, \bibinfo
  {author} {\bibfnamefont {G.~H.}\ \bibnamefont {Booth}}, \bibinfo {author}
  {\bibfnamefont {S.}~\bibnamefont {Guo}}, \bibinfo {author} {\bibfnamefont
  {Z.}~\bibnamefont {Li}}, \bibinfo {author} {\bibfnamefont {J.}~\bibnamefont
  {Liu}}, \bibinfo {author} {\bibfnamefont {J.~D.}\ \bibnamefont {McClain}},
  \bibinfo {author} {\bibfnamefont {E.~R.}\ \bibnamefont {Sayfutyarova}},
  \bibinfo {author} {\bibfnamefont {S.}~\bibnamefont {Sharma}}, \bibinfo
  {author} {\bibfnamefont {S.}~\bibnamefont {Wouters}},\ and\ \bibinfo {author}
  {\bibfnamefont {G.~K.}\ \bibnamefont {Chan}},\ }\bibfield  {title} {\bibinfo
  {title} {Pyscf: the python‐based simulations of chemistry framework},\
  }\href {https://doi.org/10.1002/wcms.1340} {\bibfield  {journal} {\bibinfo
  {journal} {Wiley Interdisciplinary Reviews: Computational Molecular Science}\
  }\textbf {\bibinfo {volume} {8}},\ \bibinfo {pages} {e1340} (\bibinfo {year}
  {2017})},\ \Eprint
  {https://arxiv.org/abs/https://onlinelibrary.wiley.com/doi/pdf/10.1002/wcms.1340}
  {https://onlinelibrary.wiley.com/doi/pdf/10.1002/wcms.1340} \BibitemShut
  {NoStop}%
\bibitem [{\citenamefont {Seeley}\ \emph {et~al.}(2012)\citenamefont {Seeley},
  \citenamefont {Richard},\ and\ \citenamefont {Love}}]{Seeley2012}%
  \BibitemOpen
  \bibfield  {author} {\bibinfo {author} {\bibfnamefont {J.~T.}\ \bibnamefont
  {Seeley}}, \bibinfo {author} {\bibfnamefont {M.~J.}\ \bibnamefont
  {Richard}},\ and\ \bibinfo {author} {\bibfnamefont {P.~J.}\ \bibnamefont
  {Love}},\ }\bibfield  {title} {\bibinfo {title} {{The Bravyi-Kitaev
  transformation for quantum computation of electronic structure}},\ }\bibfield
   {journal} {\bibinfo  {journal} {Journal of Chemical Physics}\ }\textbf
  {\bibinfo {volume} {137}},\ \href {https://doi.org/10.1063/1.4768229}
  {10.1063/1.4768229} (\bibinfo {year} {2012}),\ \Eprint
  {https://arxiv.org/abs/1208.5986} {arXiv:1208.5986} \BibitemShut {NoStop}%
\bibitem [{\citenamefont {Scuseria}(1991)}]{Scuseria1991}%
  \BibitemOpen
  \bibfield  {author} {\bibinfo {author} {\bibfnamefont {G.~E.}\ \bibnamefont
  {Scuseria}},\ }\bibfield  {title} {\bibinfo {title} {{The open-shell
  restricted Hartree-Fock singles and doubles coupled-cluster method including
  triple excitations CCSD (T): application to C$^+_3$}},\ }\href
  {https://doi.org/10.1016/0009-2614(91)90005-T} {\bibfield  {journal}
  {\bibinfo  {journal} {Chemical Physics Letters}\ }\textbf {\bibinfo {volume}
  {176}},\ \bibinfo {pages} {27} (\bibinfo {year} {1991})}\BibitemShut
  {NoStop}%
\end{thebibliography}%
\end{document}